\documentclass[aps,preprint]{revtex4}%
\usepackage{amsfonts}
\usepackage{amsmath}
\usepackage{amssymb}
\usepackage{graphicx}%
\begin{document}
\title{{\LARGE MISCONCEPTIONS ON RELATIVITY GRAVITATION AND COSMOLOGY}}
\author{Marcelo Samuel Berman }
\affiliation{Instituto Albert \ Einstein }
\affiliation{Av. Candido Hartmann, 575 - \# 17}
\affiliation{80730-440 - Curitiba - PR - Brazil - email: msberman@institutoalberteinstein.org}
\date{(Original date: June 30, 2002 - This Version date: July 26, 2008)}

\begin{abstract}
Fifteen misconceptions involving Relativity, Gravitation and Cosmology are
exposed, along with corrections.

\end{abstract}
\maketitle

\begin{center}
{\LARGE MISCONCEPTIONS ON RELATIVITY GRAVITATION AND COSMOLOGY}

\bigskip

Marcelo Samuel Berman
\end{center}

\bigskip

{\large Introduction}

\bigskip

By perusing several elementary Physics textbooks, I was astonished with the
inclusion of misconceptions dealing with Relativity, Gravitation and
Cosmology. I have to confess that most Physics textbooks published in English
can be recommended for beginner students.

{\Large The misconceptions and the corrections}

1. \textquotedblright Time is relative so there is not some master grandfather
clock that controls time in the Universe\textquotedblright.(sic)

[The comoving cosmic time is such a master clock; when I say that the Universe
is 14 billion years old, this is absolute time. A comoving observer is always
at rest relative to the matter in its neighborhood. All the literature in
Cosmology is based in such observers].

2.\textquotedblright The laws of Physics are the same for all inertial
observers\textquotedblright. (sic)

[Einstein outlined his theory with the use of tensor notation and four
dimensional space-time; only when the laws are cast in tensor notation, and in
4D spacetime, the laws of Physics retain the same form for all observers. In
three-dimensional space and common elementary mathematics, the laws of Physics
can be very awkward depending on the observer. For instance, the Maxwell
equations of electromagnetism can be very complicated for an observer moving
relative to another at rest.].

3.\textquotedblright You define relativistic 3-momentum, and then you assume
conservation when there are no forces\textquotedblright. (sic)

[In fact, the concept of force must be generalized from Newtonian Physics into
Special Relativity, so that, the relativistic force is equal to the proper
time derivative of relativistic linear momentum, but the textbooks forget to
generalize the 2nd. law. The examples in the textbooks only deal with
collisions, when the external force is supposed to be null.].

\bigskip4.\textquotedblright Today, when atomic clocks are transported from
one place to another, the time dilation of Special Relativity is to be always
taken into account.\textquotedblright\ (sic)

[Because differences in gravitational potential also affect measures of time,
the difference in gravitational potential is also to be taken into account.].

5.\textquotedblright Inertial reference systems are those where Newton%
\'{}%
s laws are valid.\textquotedblright\ (sic)

[Newton's laws are only valid for low speeds and weak gravitational fields,
but the definition of inertial systems of references is also necessary in
Special Relativity, where Newtonian laws need modification.].

6.\textquotedblright Special Relativity is not Classical
Physics.\textquotedblright\ (sic)

[Classical Physics means Non-Quantum Physics; so that, Special Relativity and
General Relativity, along with the Electromagnetism, are all Classical Theories.].

7.\textquotedblright The principle of equivalence states that gravitation and
acceleration are equivalent.\textquotedblright\ (sic)

[This is only valid at a given point and its infinitesimal neighbourhood.
There is no point in substituting a gravitational acceleration for, say, a
centripetal acceleration. Both may be numerically equal in a certain point of
space; nevertheless, the gravitational acceleration decays with the inverse
square of distance, while the centripetal one increases with distance so that,
if they are equal in one point they may be different in other locations.].

8.\textquotedblright Newtonian Physics deals with low
speeds.\textquotedblright\ (sic)

[It should be always remembered that it is only valid for weak gravitational
fields. If the speeds are not low, but the gravitational field is weak,
Special Relativity is to be taken into account; if the gravitational field is
intense, General Relativity must play a r\^{o}le].

9.\textquotedblright Edwin Hubble studied distant galaxies.\textquotedblright\ (sic)

[In fact the linear relation is valid for not too distant galaxies, which were
those he could observe in the year 1929].

10.\textquotedblright The principle of equivalence says that a homogeneous
gravitational field is completely equivalent to a uniformly accelerated
reference frame.\textquotedblright\ (sic)

[This is correct, but represents a trivial consequence of the true principle.
See \#7].

11.\textquotedblright The gravitational field is an example of a vector
field.\textquotedblright\ (sic)

[Einstein's General Theory of Relativity, is based on the assumption that the
gravitational field is represented by a metric tensor, which represents the
potentials of gravitation. This metric tensor is not a vector. ].

\bigskip

12. "With Einstein's postulates, i.e., the equivalence and covariance
principles, we find General Relativity." (sic)

[In fact, the mentioned postulates can support other theories, which have
different field equations, like for instance, scalar-tensor gravitation.]

\bigskip

13. "The Universe has a center." (sic)

[In most models, the Universe is homogeneous, so that any point is equivalent
to any other one, and may be considered the center. I call this \textit{the
egocentric observer postulate.}]

\bigskip

14. ""Dark matter" and "dark energy", are the same thing, because of
Einstein's mass-energy relation". (sic)

[In fact, dark matter and dark energy are different concepts: the first one
responds for the missing of visible matter, which should represent about 33\%
of all matter, but only 5\% is visible; 27\% is the dark matter. Dark energy,
which is 67\% of the energy density in the Universe, is there because the
Universe must have been made of critical energy density, which is the value
that turns the Universe a "flat" tri-dimensional one; the agent causing this
dark energy, responds for the acceleration of the present Universe].

\bigskip

15. "The Universe expands, so it accelerates". (sic)

[What accelerates, is the rate of expansion, and not the "center", which is at
any point].

\bigskip

{\large Conclusions}

Even good and recommendable textbooks should make an effort to bring a correct
picture also in Relativity, Gravitation and Cosmology: what we commented in
this paper should be taken as an alert on otherwise high praised books.

\bigskip

{\Large Acknowledgements}

\bigskip

\bigskip I thank Prof. M.M. Som and F.M. Gomide for conversations on the
subject of this paper. I recognize the good service of Mr. Marcelo Fermann
Guimar\~{a}es in typing this and many other manuscripts, and the encouragement
of Albert, Paula and Geni.

\end{document}